# Photonic real time video image signal processor at 17Tb/s based on a Kerr microcomb

Mengxi Tan,[1,2,3] Xingyuan Xu,[4] Andreas Boes,[3,5] Bill Corcoran,[6] Thach G. Nguyen,[4] Sai T. Chu,[7] Brent E. Little,[8] Roberto Morandotti,[9] Jiayang Wu,[2] Arnan Mitchell,[3] and David J. Moss [2*]

[1] *School of Electronic and Information Engineering, Beihang University, Beijing 100191, China.*
[2] *Optical Sciences Centre, Swinburne University of Technology, Hawthorn, VIC 3122, Australia.*
[3] *School of Engineering, RMIT University, Melbourne, VIC 3001, Australia.*
[4] *State Key Laboratory of Information Photonics and Optical Communications, Beijing University of Posts and Telecommunications, Beijing, 100876, China.*
[5] *Institute for Photonics and Advanced Sensing (IPAS) and School of Electrical and Electronic Engineering, University of Adelaide, Adelaide, 5005 SA, Australia.*
[6] *Department of Electrical and Computer System Engineering, Monash University, Clayton, VIC 3168 Australia.*
[7] *Department of Physics and Material Science, City University of Hong Kong, Tat Chee Avenue, Hong Kong, China.*
[8] *Xi'an Institute of Optics and Precision Mechanics of CAS, Xi'an, China.*
[9] *INRS-Énergie, Matériaux et Télécommunications, 1650 Boulevard Lionel-Boulet, Varennes, Québec, J3X 1S2, Canada.*
[*] *Corresponding author: dmoss@swin.edu.au*

**Signal processing has become central to many fields, from coherent optical telecommunications, where it is used to compensate signal impairments, to video image processing. Image processing is particularly important for observational astronomy, medical diagnosis, autonomous driving, big data and artificial intelligence. For these applications, signal processing traditionally has mainly been performed electronically. However these, as well as new applications, particularly those involving real time video image processing, are creating unprecedented demand for ultrahigh performance, including high bandwidth and reduced energy consumption. Here, we demonstrate a photonic signal processor operating at 17 Terabits/s and use it to process video image signals in real-time. The system processes 400,000 video signals concurrently, performing 34 functions simultaneously that are key to object edge detection, edge enhancement and motion blur. As compared with spatial-light devices used for image processing, our system is not only ultra-high speed but highly reconfigurable and programable, able to perform many different functions without any change to the physical hardware. Our approach is based on an integrated Kerr soliton crystal microcomb, and opens up new avenues for ultrafast robotic vision and machine learning.**

1. Introduction

Image processing, the application of signal processing techniques to photographs or videos, is a core part of emerging technologies such as machine learning and robotic vision [1], with many applications to LIDAR for self-driving cars [2], remote drones [3], automated in-vitro cell-growth tracking for virus and cancer analysis [4], optical neural networks [5], ultrahigh speed imaging [6, 7], holographic three-dimensional (3D) displays [8-9], and others. Many of these require real-time processing of massive real-world information, placing extremely high demands on the processing speed (bandwidth) and throughput of image processing systems. While electrical digital signal processing (DSP) technologies [10] are well established, they face intrinsic limitations in energy consumption and processing speed such as the well-known von Neumann bottleneck [11].

To overcome these limitations, optical signal processing offers the potential for much higher speeds [2], and this has been achieved using a variety of techniques including silicon photonic crystal metasurfaces [12], surface plasmonic structures [13], and topological interfaces [14]. These free-space, spatial-light devices offer many attractions such as compact footprint, low power consumption, and compatibility with commercial cameras and optical microscopes. However, they tend to be non-reconfigurable fixed systems designed to perform a single fixed function. On a more advanced level, human action recognition through processing of video image data, has been achieved using photonic computers [15, 16]. However, these were achieved either in comparatively low speed systems [15] or in high bandwidth (multi-TeraOP regime) systems based on bulk-optics that is incompatible with integration [16]. To date, optical systems, especially those compatible with integration [17], still have not demonstrated that are capable of processing of large data sets of high-definition video images and at ultra-high speeds – enough for real-time video image processing.

Here, we demonstrate an optical real-time signal processor for video images that is reconfigurable and compatible with integration. It is based on components that are either already integrated or have been demonstrated in integrated form, and operates at an ultra-high bandwidth of 17 Terabits/s. This is sufficient to process approximately 400,000 (399,061) video signals both concurrently and in real-time, performing up to 34 functions on each signal simultaneously.

Here, the term "functions" refers to signal processing operations comprised of fundamental mathematical operations that are performed by the system, which in our case relate to object image edge enhancement, detection and motion blur. These functions operate on the input signal to extract or enhance these key characteristics and include both integral and fractional order differentiation, fractional order Hilbert transforms, and integration. For differentiation and Hilbert transforms we perform both integral order and a continuous range of fractional order transforms. Therefore, while there are 3 basic types of functions that we perform, with the inclusion of a range of integral and fractional orders we achieve 34 functions in total. Importantly, these 34 functions are all achieved without any change in hardware, but only by tuning the parameters of the system. Furthermore, beyond these 34 functions, the range of possible functions is in fact unlimited given that the system can process a continuous range of arbitrary fractional and high-order differentiation and fractional Hilbert transforms.

Our system is comparable to electrical digital signal processing (DSP) systems but with the important advantages that it operates at multi-terabit/s speeds, enabled by massively parallel processing. It is also very general, flexible, and highly reconfigurable – able to perform a wide range of functions without requiring any change in hardware. We perform multiple image processing functions in real-time, which are essential for machine vision and microscopy for tasks such as object recognition or identification, feature capture, and data compression [12-13]. We use an integrated Kerr soliton crystal microcomb source that generates 95 discrete taps, or wavelengths as the basis for massively parallel processing, with single channel rates at 64 GigaBaud (pixels/s). Our experimental results agree well with theory, demonstrating that the processor is a powerful approach for ultrahigh-speed video image processing for robotic vision, machine learning, and many other emerging applications.

## 2. Results

**a) Principle of Operation**

The operational principle of the video image processor is based on the RF photonic transversal filter [18-20] approach, as represented by Equation (1) and illustrated in Figure 1(a), (b), (c-i), (c-ii), (c-iii). We employ wavelength division multiplexed (WDM) signals to provide the different taps, or channels– each wavelength representing a single tap/channel. We also use WDM as a central means of accomplishing both single and multiple functions simultaneously. The tap delays required by Equation 1 are achieved here by an optical delay line in the form of standard single mode fibre (SMF) in order to perform the wavelength (ie., tap or channel) dependent delays. We use WaveShaper to flatten the comb and implement the channel weightings for the transversal filter for each function, as well as to separate different groups of wavelengths to perform parallel and simultaneous processing of multiple functions. We used a maximum of 95 wavelengths supplied by a soliton crystal Kerr microcomb that produced a comb spacing of about 50 GHz. The transfer function of the system is given by

$$H(\omega) = \sum_{n=0}^{N-1} h(n) e^{-j\omega nT} \quad (1)$$

where $\omega$ is the RF angular frequency, $T$ is the time delay between adjacent taps (i.e., wavelength channels), and $h(n)$ is the tap coefficient of the $n^{\text{th}}$ tap, or wavelength, which can be calculated by performing the inverse Fourier transform of $H(\omega)$ [18-20]. In Eq. (1), the tap coefficients can be tailored by shaping the power of comb lines according to the different computing functions (e.g., differentiation, integration, and Hilbert transformation), thus enabling different video image processing functions. For microcombs with multiple equally spaced comb lines transmitted over the dispersive SMF, in Eq. (1) $T$ is given by $T = D \times L \times \Delta\lambda$, where $D$ is the dispersion coefficient of the SMF, $L$ is the length of the SMF, $\Delta\lambda$ is the spectral separation between adjacent comb lines (in our case 48.9 GHz) and the RF bandwidth of the system is given by $f = 1/T$. The optical delay lines play a crucial role in achieving simultaneous processing by introducing wavelength dependent controlled time delays to the different channels, enabling the functions to be processed independently and in parallel. These time delays coincide with the requirements of the transversal filter function (Equation 1) and ensure that the input signals for each function are properly aligned and synchronized. To change the system bandwidth one needs to change the time delay between adjacent wavelengths. While this is generally fixed for a given system, it can be changed by either using different lengths of SMF or alternatively adding a length of dispersion compensating fibre (DCF) which effectively reduces the net dispersion $D$ of the fibre, equivalent to decreasing the SMF fibre length. To achieve dynamic tuning of the RF bandwidth would require a tunable delay line which is beyond the scope of our work.

All 95 wavelengths from the microcomb were passed through a single output port WaveShaper which flattened the comb and weighted the individual lines according to the required tap weights for the particular function being performed. The weighted wavelengths were then passed through an electro-optic modulator which was driven by the analog input video signal. The output function was finally generated by summing all of the wavelengths, achieved by photodetection of all wavelengths.

The setup for the massively parallel signal processing demonstration is shown in Figure 2, which uses an approach similar to that used for our ultrahigh speed optical convolution accelerator [5]. Figure 2 shows the results for 34 different functions, which are listed in detail along with their individual parameters in Supp. Table S1. In this work, for the ultrahigh speed demonstration we chose fewer taps for each function – typically 5 – in order to increase the number of functions we could perform, while maximizing the overall speed or bandwidth. We found that 5 taps was the minimum number that was able to achieve good performance, striking a balance between complexity and efficiency, providing satisfactory performance for the desired functions while minimizing the number of required elements. We were able to achieve 34 functions simultaneously overall. As for the initial tests, the microcomb lines were fed into a WaveShaper then weighted and directed to an EO modulator followed by the SMF delay fibre. Lastly, the generated RF signal for each function was resampled and converted back to digital video image frames, which formed the digital output signal of the system.

In our experiments the analog input video image frames were first digitized and then flattened into 1D vectors (**X**) and encoded as the intensities of temporal symbols in a serial electrical waveform by a high speed analog to digital converter with a resolution of 8 bits per symbol at a sampling rate of 64 gigabaud (see Methods). In principle, for analog video signals this A/D and D/A step can be avoided. We added this step since this allowed us to dramatically increase the speed of the video signal over standard video rates in order to fully exploit the ultrahigh speed of our processor.

The use of WDM and WaveShapers allows for very flexible allocation of wavelengths and highly reconfigurable tuning. By employing these components in a carefully designed configuration where each function only requires a limited number of wavelengths, multiple functions can be simultaneously processed. By configuring the WaveShapers appropriately according to the transfer function, different functions can be applied to different groups of wavelength channels, facilitating simultaneous processing of multiple functions across the entire microcomb spectrum.

In the experiments presented here, we demonstrate real-time video image processing, simultaneously executing 34 functions encompassing edge enhancement, edge detection, and motion blur. Edge detection serves as the foundation for object detection, feature capture, and data compression [12-13]. We achieve this by temporal signal differentiation with either high integral or fractional order derivatives that extract information about object boundaries within images or videos. We also perform a motion blur function based on signal integration that represents the apparent streaking of moving objects in images or videos. It usually occurs when the image being recorded changes during the recording of a single exposure, and has wide applications in computer animation and graphics [51]. Edge enhancement or sharpening based on signal

Hilbert transformation is also a fundamental processing function with wide applications [52]. It enhances the edge contrast of images or videos, thus improving their acutance. The standard Hilbert transform implements a 90 degree phase shift and is commonly used in signal processing to generate a complex analytic signal from a real-valued signal. We also employ arbitrary, or fractional order, Hilbert transforms which have been shown to be particularly useful for object image edge enhancement [40]. These processing functions not only underpin conventional image or video processing [53-54] but also facilitate emerging technologies such as robotic vision and machine learning [2, 4].

To achieve fractional order operations, the system utilizes the concept of optical fractional differentiation and Hilbert transforms [45]. This is accomplished by carefully designing the WaveShapers in the optical signal processing setup with the appropriate set of weights (phase and amplitude) for shaping the optical signals in the frequency domain, and their configurations can be adjusted to achieve different fractional order operations. While we use off-the-shelf commercial WaveShapers, in practice these can be realized using various techniques compatible with integration, such as cascaded Mach-Zehnder interferometers or programmable phase modulators. These components enable precise control over the spectral phase and amplitude profiles of the optical signals, allowing the realization of fractional order operations.

A key feature of our system that was critical in achieving high fidelity and performance signal processing was the improvement we obtained in the frequency comb spectral line shaping accuracy. To accomplish this we employed a two-stage shaping strategy (see methods) [55] where a feedback control path was employed to calibrate the system and further improve the comb shaping accuracy. The feedback loop in the optical signal processing system plays a crucial role in ensuring the optimization of signal processing quality. It involves monitoring the system's output and feeding it back to adjust the tap weights in order to achieve the best performance (see Methods for more detail). The error signal was generated by directly measuring the impulse response of the system and then comparing it with ideal tap coefficients. Note that this type of feedback calibration approach is challenging and rarely used for analog optical signal computing, such as the systems based on either spatial-light metasurfaces [12, 56] or waveguide resonators [35, 57], for example.

Our system is based on a soliton crystal (SC) microcomb source, generated in an integrated MRR [18-20] (Figure 3(a), (b), Supplementary Note 1, Supplementary Figures S1-S7). Since their first demonstration in 2007 [21], and subsequently in CMOS compatible integrated form [22], optical frequency combs generated by compact micro-scale resonators, or micro-combs [22-25], have led to significant breakthroughs in many fields such as metrology [26], spectroscopy [25, 27], telecommunications [23, 28], quantum optics [29-30], and radio-frequency (RF) photonics [18-20, 31-34]. Microcombs offer new possibilities for major advances in emerging applications such as optical neural networks [5], frequency synthesis [35], and light detection & ranging (LIDAR) [2, 36-37]. With a good balance between gain and cavity loss as well as dispersion and nonlinearity, soliton microcombs feature high coherence and low phase noise and have been highly successful for many RF photonic applications [19, 38-43].

SC microcombs, multiple self-organized solitons [44], have been highly successful particularly for RF photonic signal processing [18-20, 45-49], ultra-dense optical data transmission [23], and optical neuromorphic processing [5, 50]. They feature very high coherence with low phase noise [49], are intrinsically stable with only open-loop control (Supplementary Note 1, Supplementary Figure S2 and Supplementary Movie S1) and can be simply and reliably initiated via manual pump wavelength sweeping. Further, they have intrinsically high conversion efficiency since the intra-cavity energy is much higher than for single soliton states [5, 23]. Our microcomb has a low free spectral range (FSR) of ~48.9 GHz, closely matching the ITU frequency grid of 50GHz, and generates over 80 Iwavelengths in the telecom C-band, which serves as discrete taps for the video image processing system.

**b) Experimental Results – Static Images**

Since each frame of a streaming video signal is essentially a static image, we initially benchmarked the system single function system performance on static images with varying numbers of taps to understand the tradeoffs in performance. Figure 4 (a-i) (a-ii) (a-iii) (b-i) (b-ii) (b-iii) (c-i) (c-ii) (c-iii) (d-i) (d-ii) (d-iii) (e-i) (e-ii) (e-iii) (f-i) (f-ii) (f-iii) (g-i) (g-ii) (g-iii) (h-i) (h-ii) (h-iii) (l-i) (l-ii) (l-iii) shows the experimental results of static image processing using the above RF photonic system. We conducted initial experiments to investigate the performance of the transfer functions with 15, 45, and 75 taps for single function performance, where the WaveShaper was set to zero out any unneeded wavelengths. These experiments were performed on single static images – ie., a single frame of the video signal, as shown in Figure 4. This was aimed at exploring the influence of the tap number on the signal processing performance and determining the optimal tap number for the subsequent massively parallel signal processing demonstration. This allowed us to assess these trade-offs between complexity, accuracy, and efficiency in the signal processing operations.

The original (unprocessed) high definition (HD) digital images had a resolution of 1080 × 1620 pixels. The results for edge detection based on signal differentiation with orders of 0.5, 0.75, and 1 are shown in Figures 2(a) – (c), respectively. In each figure, we show (i) the designed and measured spectra of the shaped comb, (ii) the measured and simulated spectral response, and (iii) the measured and simulated images after processing. The measured comb spectra and spectral response were recorded by an optical spectrum analyser (OSA) and a vector network analyser (VNA), respectively. The experimental results agree well with theory, indicating successful edge detection for the original images.

In Figs. 3(d) – (f), we show the results for motion blur based on signal integration with different tap numbers of 15, 45, and 75, respectively. These are also in good agreement between the experimental results and theory. The blur intensity increases with the increased number of taps, reflecting the fact that there is improved processing performance as the number of taps increases. Compared with discrete laser arrays that feature bulky sizes, limiting the number of available taps, microcombs generated by a single MRR can operate as a multi-wavelength source that provides a large number of wavelength channels, as well as greatly reducing the size, power consumption, and complexity. This is very attractive for the RF photonic transversal filter system that requires a large number of taps for improved processing performance.

Figs. 3(g) – (I) show the results of edge enhancement based on signal Hilbert transformation (90° phase shift) with different operation bandwidths of 12 GHz, 18 GHz, and 38 GHz, respectively. In our experiment, the operation bandwidth was adjusted by changing the comb spacing (2 FSRs vs 3 FSRs of the MRR) and the fibre length (1.838 km vs 3.96 km). Note that having to change the fibre length in principle can be avoided by using tunable dispersion compensators [58-59]. The WaveShaper can accommodate any FSR (channel spacing) as long as it fits roughly within telecom band channel spacings. Thus the system is very flexible and can accommodate any FSR by the WaveShaper or delay by changing the length of SMF if a different microcomb device is used. Alternatively, the WaveShaper can be used to filter out certain channels if a larger effective channel spacing is desired compared to the source FSR. The tradeoff in varying the FSR is that smaller FSRs yield lower bandwidths whereas larger FSRs reduce the number of wavelengths within the telecom C band. As can be seen, the edges in the images are enhanced, and the experimental results are consistent with the simulations.

We also demonstrate more specific image processing such as edge detection based on fractional differentiation with different orders of 0.1 – 0.9, edge enhancement based on fractional Hilbert transformation with different phase shifts of 15° – 75°, and edge detection with different operation bandwidths of 4.6 GHz – 36.6 GHz (Supplementary Note 2, Supplementary Figures S3 – S6). By changing the relevant parameters, this resulted in processed images with different degrees of edge detection, motion blur, and edge enhancement. By simply programming the WaveShaper to shape the comb lines according to the designed tap coefficients, different image processing functions were realized without changing the physical hardware. This reflects the high reconfigurability of our video image processing system, which is challenging for image processing based on spatial-light devices [12-14]. In practical image processing, there is not one single processing function that has one set of parameters that can meet all the requirements. Rather, each processing function requires its own unique set of tap weights. Hence, the high degree of reconfigurability and versatility of our image processing system is critical to meet diverse and practical processing requirements.

**c) Experimental Results – Real-Time Video**

In addition to static image processing, our microcomb based RF photonic system can also process dynamic videos in real-time. Our results for real-time video processing are provided in supplementary Movie S2, while Supplementary Figures S8-S10 show samples of these experimental results. The supplementary Movie S2. starts off with the first original source video frames and is followed by the simulation and experiment results shown side by side for the differentiator, integrator, and Hilbert transformer. This is then followed by the 34 functions (Supplementary Note 2, Supplementary Table S1) performed simultaneous by the massively parallel video processor. Finally, high order of derivatives are shown, and the video ends with results based on full 2 dimensional derivatives (see below).

The first original video had a resolution of 568 × 320 pixels and a frame rate of 30 frames per second. Supplementary Figure S8(a) shows 5 frames of the original video, together with the corresponding electrical waveform after digital-to-analogue conversion. Supplementary Figures S8(b) – (d) show the corresponding results for the processed video after edge detection (0.5-order fractional differentiation), motion blur (integration with 75 taps), and edge enhancement based on a Hilbert transformation with an operation bandwidth of 18 GHz, respectively. As for the static image processing, the real-time video processing results show good agreement with theoretical predictions.

To fully exploit the bandwidth advantage of optical processing, we further performed massively parallel real-time multi-functional video processing. The experimental setup and results are shown in Figure 4 and Supplementary Note 2, Supplementary Figures S8-S10. We used 95 comb lines around the C band in our demonstration. After flattening and splitting the comb lines via the first WaveShaper, we obtained 34 parallel processors, most of which consisted of five taps. We simultaneously performed 34 video image processing functions, including fractional differentiation with fractional order from 0.05 to 1.1, fractional Hilbert transformation with phase shift from 65° to 90°, an integrator, and bandpass Hilbert transformation with a 90° phase shift (see Supplementary Table S1 for detailed parameters for each function). The corresponding total processing bandwidth equals 64 GBaud × 34 (functions) × 8 bits = 17.4 Terabit/s – well beyond the processing bandwidth of electrical video image processors [10].

3. Discussion

To analyze the performance of our video image processor, we evaluated the processed images based on the ground truth for both quantitative and qualitative comparisons [60]. We used respective ground truths for the evaluation of 3 BSD (Berkeley Segmentation Database) images after edge detection and compared relevant performance parameters with the same images processed based on the widely used Sobel's algorithm [61]. (In signal processing and data analysis, "Ground Truth" typically refers to the objectively true or correct values or information that serves as a reference for evaluating the performance or accuracy of a system or algorithm.) Figure 5 shows the images processed using Sobel's algorithm and our video image processor (including differentiation with different orders from 0.2 to 1.0). The comparison of the performance parameters including performance ratio (PR) and F-Measure is provided in Table 1, where higher values of these parameters reflect a better edge detection performance. As can be seen, our differentiation results for PR and F-Measures are better than Sobel's approach, reflecting the high performance of our video image processor.

The maximum input rate we used was 64 GBaud, or Gigapixels/s. This, combined with the fact that we performed 34 channels with a video resolution of 568 × 320 that resulted in 181,760 pixels at a frame rate of 30 Hz, yields 5,452,800 pixels/s, resulting in simultaneous real-time processing of $64 \times 10^9 \times 34/ (181,760 \times 30)$ = 399,061 video signals per second. For HD videos (720 × 1280 = 921,600 pixels) at a frame rate of 50 Hz, this equates to about 47,222 video signals in parallel. The processing throughput can be increased even further by using more comb lines in the L-band.

We provide the root mean square errors (RMSEs) in Supplementary Figure S7 and Table S2 to quantitatively assess the agreement between the measured waveforms and the theoretical results for different image processing functions. We find that for the Hilbert transformer, for example, with tunable phase shift, the RMSE values ranged from 0.0586 to 0.1045, depending on the specific phase shift angle. These RMSE values provide a quantitative measure of the agreement between the experimental measurements and the theoretical predictions. Lower RMSE values generally indicate a better correspondence between the two. Our results for the RSMEs indicate that the measured waveforms closely align with the expected behavior of the respective image processing functions.

The processing accuracy of our system is lower than electrical DSP image processing but higher than analog image processing based on passive optical filters [13, 52, 57] (see Supplementary Figure S7 and Table S1). Different lengths of fibre were used to be compatible with the different spacings of the different FSRs used (set by the Waveshaper) and to achieve an optimum RF bandwidth. This is mainly a result of the hybrid nature of our system, which is equivalent to electrical DSP systems but implemented by photonic hardware. There are a number of factors that can lead to tap errors during the comb shaping, thus leading to a non-ideal frequency response of the system as well as deviations between the experimental results and theory. These mainly include a limited number of available taps, the instability of the optical microcomb, the accuracy of the WaveShapers, the gain variation with wavelength of the optical amplifiers, the chirp induced by the optical modulator, the second-order dispersion (SOD) induced power fading, and the third-order dispersion (TOD) of the dispersive fibre. Chirp-induced errors refers to distortions that arise in the signal processing system due to the presence of chirp in the optical signals. Chirp (frequency modulation or shift in time) can be caused by various factors, such as dispersion or nonlinearity in the optical components.

We encode the image pixels directly on to the optical signal using the intensity modulator. The reason we slice the input image is because our AWG performs 1 dimensional signal operations, otherwise this is not necessary. There are a variety of ways to slice the input image. The video signal is encoded without any time delay onto the optical wavelengths. For processing the signal, although SMF is used to achieve the incremental delay lines required by the transversal filter transfer function, it does not slow down the speed of the system but only adds to the latency. For the same image / video signal with the same delay line, we only need one modulator to encode the input signal. We pre- post- the image using the arbitrary waveform generator to digitize the analog signal and convert it into a high speed analog signal to enable us to perform with the full capability of our signal processor. In principle this pre- post- processing is not necessary since the system can process and output analog signals directly. The AWG does not form a fundamental part of our processor.

In terms of the energy efficiency of the optical signal processor, we use the same approach as reported elsewhere. [5] The power consumption of the comb source is estimated to be 1500 mW while that of the EDFAs is estimated at 2000 mW (100 mW for each EDFA) and for the intensity modulator is approximately 3.4 V × 0.01 A = 34 mW. The overall computing speed of the optical signal processor is 2 × 34 × 5 × 62.9 = 21.386 TeraOPs/s. As such, the energy per bit of the optical signal processor is roughly (1500 + 2000 + 34 × 19) mW / 21.386 TeraOPs/s = 0.194pJ/operation.

The number of available taps can be increased by using MRRs with smaller FSRs or optical amplifiers with broader operation bandwidths. The accuracy realized by the WaveShapers and the optical amplifiers was significantly improved by using a two-stage comb shaping strategy as well as the feedback loop calibration mentioned previously. [55] By using low-chirp modulators, the chirp-induced tap errors can be suppressed. The discrepancies induced by the SOD and TOD of the dispersive fibres can also be reduced by using a second WaveShaper to compensate for the group delay ripple of the system (see Supplementary Note 3, Supplementary Figure S11).

Our massively parallel photonic video image processor, which operates on the principle of time-wavelength-spatial multiplexing, similar to the optical vector convolutional accelerator in our previous work [5], is also capable of performing convolution operations for deep learning neural networks. This opens up new opportunities for image or video processing applications in robotic vision and machine learning. In particular, each parallel function can be trained and performed as many as 34 kernels with a size of 5 by 1 for the convolutional neural network, therefore could ultimately achieve a neural network, avoiding the bandwidth limitation given by the analogue-to-digital converters. Note that, without the use of an AWG or OSC, our processor could directly process analog signals, while with the use of an AWG and OSC it can also process digital signals. Hence it effectively is equivalent to electrical DSP.

Although the system presented here operated at a speed of approximately 17 Terabits/s, it is highly scalable in speed. Figure S12 shows a video image processor using the C + L + S bands (with 405 wavelengths distributed over 81 processors with 5 taps each in size) and 19 spatial paths, all exploiting polarization, yielding a total processing bandwidth of 1.575 Petabits/s (Supplementary Note 4).

Our system is highly compatible with integrated technologies and so there is a strong potential for much higher levels of integration, even reaching full monolithic integration. The core component of our system, the microcomb, is already fully integrated. Further, all of the other components have been demonstrated in integrated form, including integrated InP spectral shapers [59], high-speed integrated lithium niobite modulators [63], integrated dispersive elements [59], and photodetectors [65]. Finally, low power consumption and highly efficient microcombs have been demonstrated with single soliton states [66] and laser cavity-soliton Kerr combs [67-68], which would greatly reduce the energy requirements. A key advantage to monolithic integration would be the ability to integrate electronic elements on-chip such as an FPGA module for feedback control. Finally, being much more compact, the monolithically integrated system should be much less susceptible to the environment, thus reducing the required level of feedback control.

## 4. Conclusions

In conclusion, we report the first demonstration of video image processing based on Kerr microcombs. Our RF photonic processing system, with an ultrahigh processing bandwidth of 17.4 Tbs/s, can simultaneously process over 399,061 video signals in real-time. The system is highly reconfigurable via programmable control, and can perform different processing functions without changing the physical hardware. We experimentally demonstrate different video image processing functions including edge detection, motion blur, and edge enhancement. The experimental results agree well with theory, verifying the effectiveness of using Kerr microcombs for ultrahigh-speed video image processing. Our results represent a significant advancement for fundamental photonic computing, paving the way for practical ultrahigh bandwidth real-time photonic video image processing on a chip.

## Methods

### 1. Microcomb generation

We use SC microcombs generated by an integrated MRR (Figure 3 and Supplementary Figures S1-S6) for video image processing. The SC microcombs, which include multiple self-organized solitons confined within the MRR, were also used for our previous demonstrations of RF photonic signal processing [45-49], ultra-dense optical data transmission [23], and optical neuromorphic processing [5, 50].

The MRR used to generate SC microcombs (Figure 3(b)) was fabricated based on a complementary metal–oxide–semiconductor (CMOS) compatible doped silica glass platform [22-23]. It has a radius of ~592 µm, a high quality factor of ~1.5 million, and a free spectral range (FSR) of ~0.393 nm (i.e., ~48.9 GHz). The low FSR results in a large number of wavelength channels, which are used as discrete taps in our RF photonic transversal filter system for video image processing. The cross-section of the waveguide was 3 µm × 2 µm, resulting in anomalous dispersion in the C-band (Supplementary Figure S1). The input and output ports of the MRR were coupled to a fibre array via specially designed mode converters, yielding a low fibre-chip coupling loss of 0.5 dB/facet.

In our experiment, a continuous-wave (CW) pump light was amplified to 30.5 dBm and the wavelength was swept from blue to red. When the detuning between pump wavelength and MRR's cold resonance became small enough, the intra-cavity power reached a threshold, and optical parametric oscillation driven by modulation instability (MI) was initiated. Primary combs (Figure 3 (d-ii) and (d-iii)) were first generated, with the comb spacing determined by the MI gain peak [23,24, 69]. As the detuning changed further, a second jump in the intra-cavity power was observed, where distinctive 'fingerprint' SC comb spectra (Figure 3 (d-iv)) appeared, with a comb spacing equal to the MRR's FSR. The SC microcomb arising from spectral interference between the tightly packaged solitons circulating along the ring cavity exhibits high coherence and low RF intensity noise (Figure 3(c)), which are consistent with our simulations (Supplementary Movie S1). It is also worth mentioning that the SC microcomb is highly stable with only open-loop temperature control (Supplementary Figure S2). In addition, it can be generated through manual adiabatic pump wavelength sweeping – a simple and reliable initiation process that also results in much higher energy conversion efficiency than single-soliton states [5].

### 2. Microcomb shaping

To achieve the designed tap weights, the generated SC microcomb was shaped in power using liquid crystal on silicon (LCOS) based spectral WaveShapers. We used two-stage comb shaping in the video image processing experiments. The generated SC microcomb was pre-flattened and split by the first WaveShaper (Finisar 16000S), which yields an improved optical signal-to-noise ratio (OSNR) and a reduced loss control range for the second-stage comb shaping. The pre-flattened and split comb was then accurately shaped by the second WaveShaper (Finisar 4000S) according to the designed tap coefficients for different video image processing functions. The positive and negative tap coefficients were achieved by separating the wavelength channels into two spatial outputs of the second WaveShaper and then detected by a balanced photodetector (Finisar BPDV2150R).

In order to improve the comb shaping accuracy, a feedback control loop was employed for the second WaveShaper. First, we used RF Gaussian pulses as the system input and measured replicas of the input pulses in different wavelength channels. Next, we extracted peak intensities of the system impulse response and obtained accurate RF-to-RF tap coefficients. Finally, the extracted tap coefficients were subtracted from the ideal tap coefficients to obtain an error signal, which was used to calibrate the loss of the second WaveShaper. After several iterations of the comb shaping loop, an accurate impulse response that compensated for the non-ideal impulse response of the system was obtained, thus significantly improving the accuracy of the RF photonic video image processing. Directly measuring the system impulse response is more accurate compared to measuring the optical power of the comb lines, given the slight difference between the two ports into the balanced detector. The shaped impulse responses for different image processing functions are shown in the Supplementary Figs. S3 – S6.

### 3. Derivative (from fractional to high order)

The transfer function of a differentiator is given by

$$H(\omega) \propto (j\omega)^N \tag{2}$$

where $j$ equals to $\sqrt{-1}$, $\omega$ represents the angular frequency, and $N$ is the order of differentiation, which in our case can be both fractional [70] and integral [18], even complex. The experiment results for both fractional and integral order differentiation can be seen in Supplementary Movie S2. The fractional-order is tunable from 0.05 to 1.1, with a step of 0.05. We achieved high order differentiation with an order of 2, 2.5, 3, which to the best of our knowledge, is the highest order of derivative that can be achieved for video image processing.

### 4. Two-dimensional video image processing

Normally, processing functions such as differentiation, operating on video signals, only result in a one dimensional process – since it acts on individual lines of the video raster image. However, by appropriately pre-processing the video signal it is possible to obtain a fully two-dimensional derivative [71]. $f_z(x, y)$ represent the $z_{th}$ frame of a video signal with $O \times P$ pixels, where $x = 0, 1, 2, \ldots, O - 1, y = 0, 1, 2, \ldots, P - 1$. Thus, the two-dimensional derivative result is given by:

$$D_z(u, v) = \sum_{u=0}^{M-1} h_x(u) e^{-j\omega_x uT} \sum_{v=0}^{N-1} h_y(v) e^{-j\omega_y vT} \tag{3}$$

where $M$, $N$ is the number of taps, $u = 0, 1, 2, \ldots, M – 1, v = 0, 1, 2, \ldots, N - 1$.

The electrical input data was temporally encoded by an arbitrary waveform generator (Keysight M8195A). The raw input matrices were first sliced horizontally and vertically into multiple rows and columns, respectively, which were flattened into vectors and connected head-to-tail. After that, the generated vectors were multicast onto different wavelength channels via a 40-GHz intensity eletro-optic modulator (iXblue). For the video with a resolution of 303× 262 pixels and a frame rate of 30 frames per second, we used a sampling rate of 64 Giga samples/s to form the input symbols. A dispersive fibre was employed to provide a progressive delay $T$. Next, the electrical output waveform was resampled and digitized by a high-speed oscilloscope (Keysight DSOZ504A) to generate the final output. The magnitude and phase responses of the RF photonic video image processing system were characterized by a vector network analyser (Agilent MS4644B 40 GHz bandwidth) working in the S21 mode. Finally, we restored the processed video into the original size of the matrix and took the average of horizontally and vertically processed video and formed this into a two-dimensional processed video (Supplementary Movie S2).

### 5. Details of the video image dataset

The high definition (HD) image with a resolution of 1080 × 1620 pixels we performed is a photo taken by Nikon D5600 in front of the Exhibition building in the centre of Melbourne city, Australia, in 2020. The video of 568 × 320 pixels was captured by a Drone Quadcopter UAV with Optical Zoom camera (DJL Mavic Air2 Zoom), this was a short trip during the eastern holiday, in 2019. The author and her friend were started from Melbourne to Adelaide, passing the pink lake and playing guitar, this was a great memory before the pandemic and continuous lockdown in Melbourne. The short video of the skateboard with a resolution of 303× 262 pixels was taken by the author using iPhone SE in front of Victoria Library, Melbourne, Australia, in 2020.

**Acknowledgements**
This work was supported by the Australian Research Council Discovery Projects Program (No. DP150104327, No. DP190101576). R. M. acknowledges support by the Natural Sciences and Engineering Research Council of Canada (NSERC) through the Strategic, Discovery and Acceleration Grants Schemes, by the MESI PSR-SIIRI Initiative in Quebec, and by the Canada Research Chair Program. Brent E. Little was supported by the Strategic Priority Research Program of the Chinese Academy of Sciences, Grant No. XDB24030000.

**Author contributions**
M. T., X. X., and D. J. M. developed the original concept. B. E. L. and S. T. C. designed and fabricated the integrated devices. M. T. performed the experiments. D. J. M., M. T., J. W., A.B., B.C., T.G.N. R.M. A.M., and X. X. contributed to the development of the experiment and to the data analysis. D. J. M., M. T., J. W., A.B., B.C., T.G.N. R.M. A.M., and X. X. contributed to the writing of the manuscript. D. J. M., X. X., J. W., and A. M. supervised the research.

**Conflict of Interest**
The authors declare that there is no conflict of interest.

**Data Availability**
All data is available upon reasonable request to the authors.

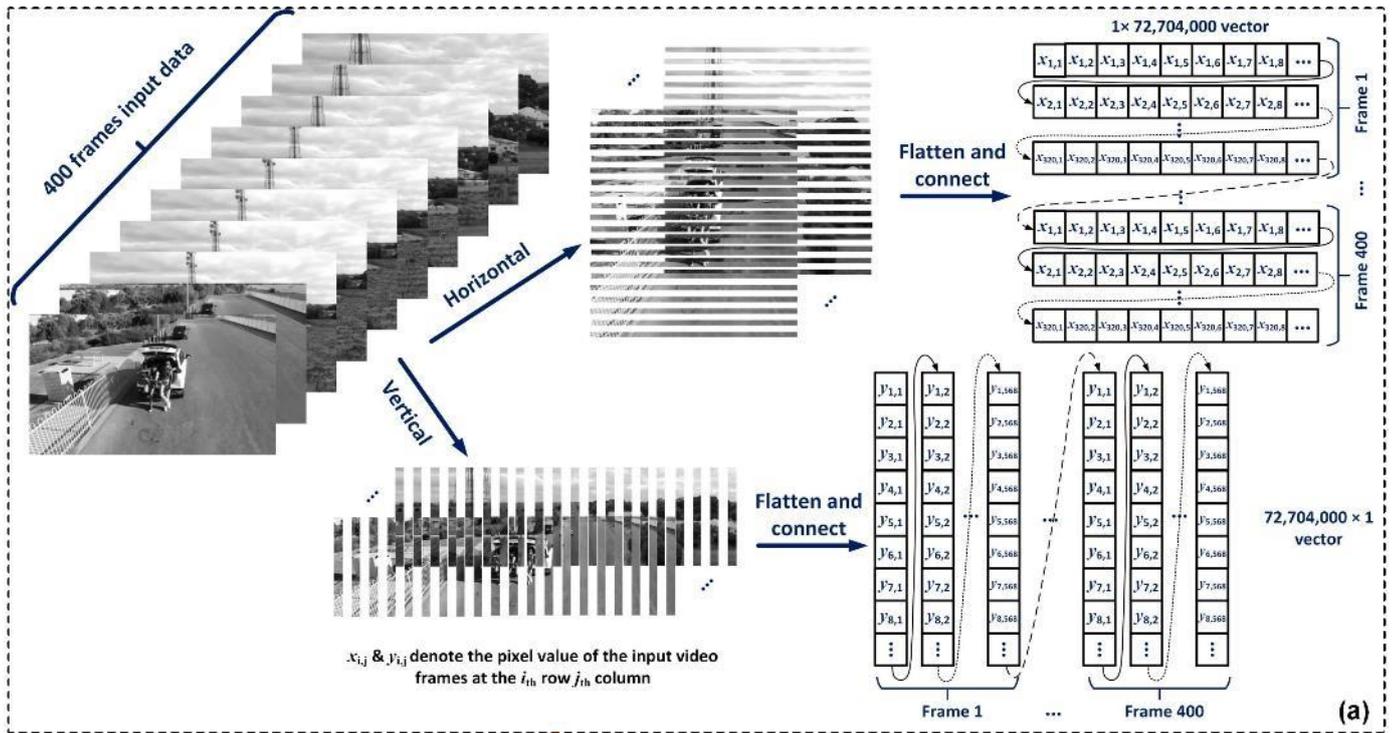
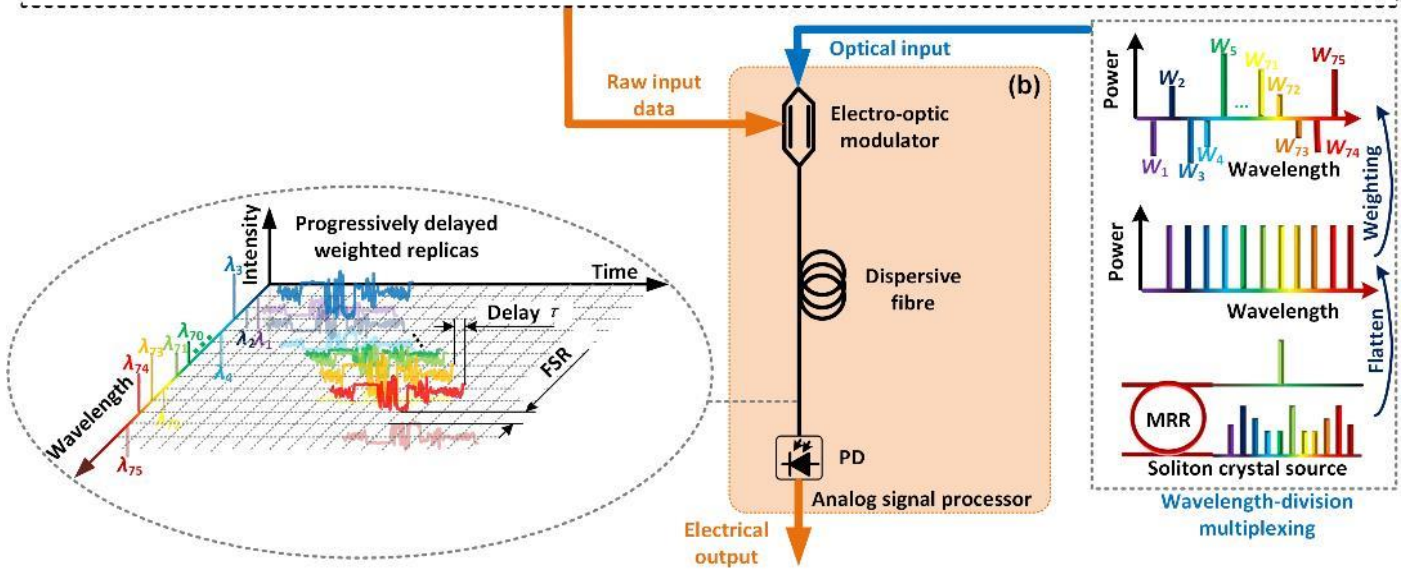
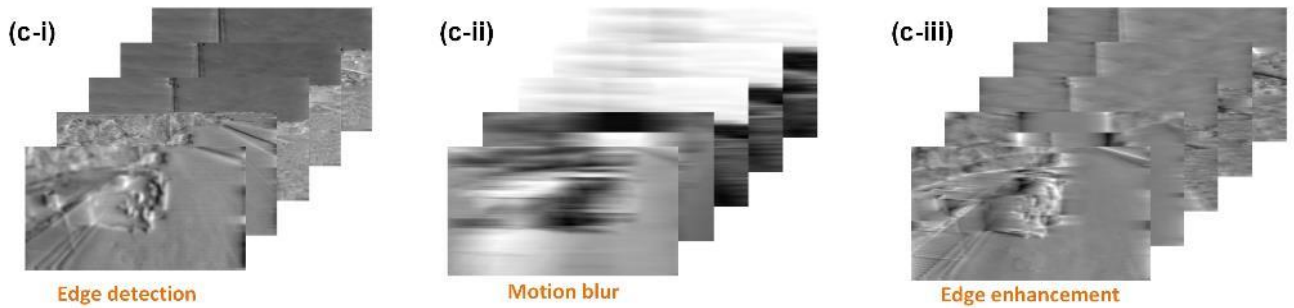

**Figure 1. Operation principle of the RF photonic video image processor.** PD: photodetector. (a) Diagram illustration of the flattening method applied to the input video frames including both horizontally and vertically. (b) Schematic illustration of experimental setup for video image processing. (c) The processed video frames after (i) 0.5 order differentiation for edge detection, (ii) integration for motion blue, and (iii) Hilbert transformation for edge enhancement.

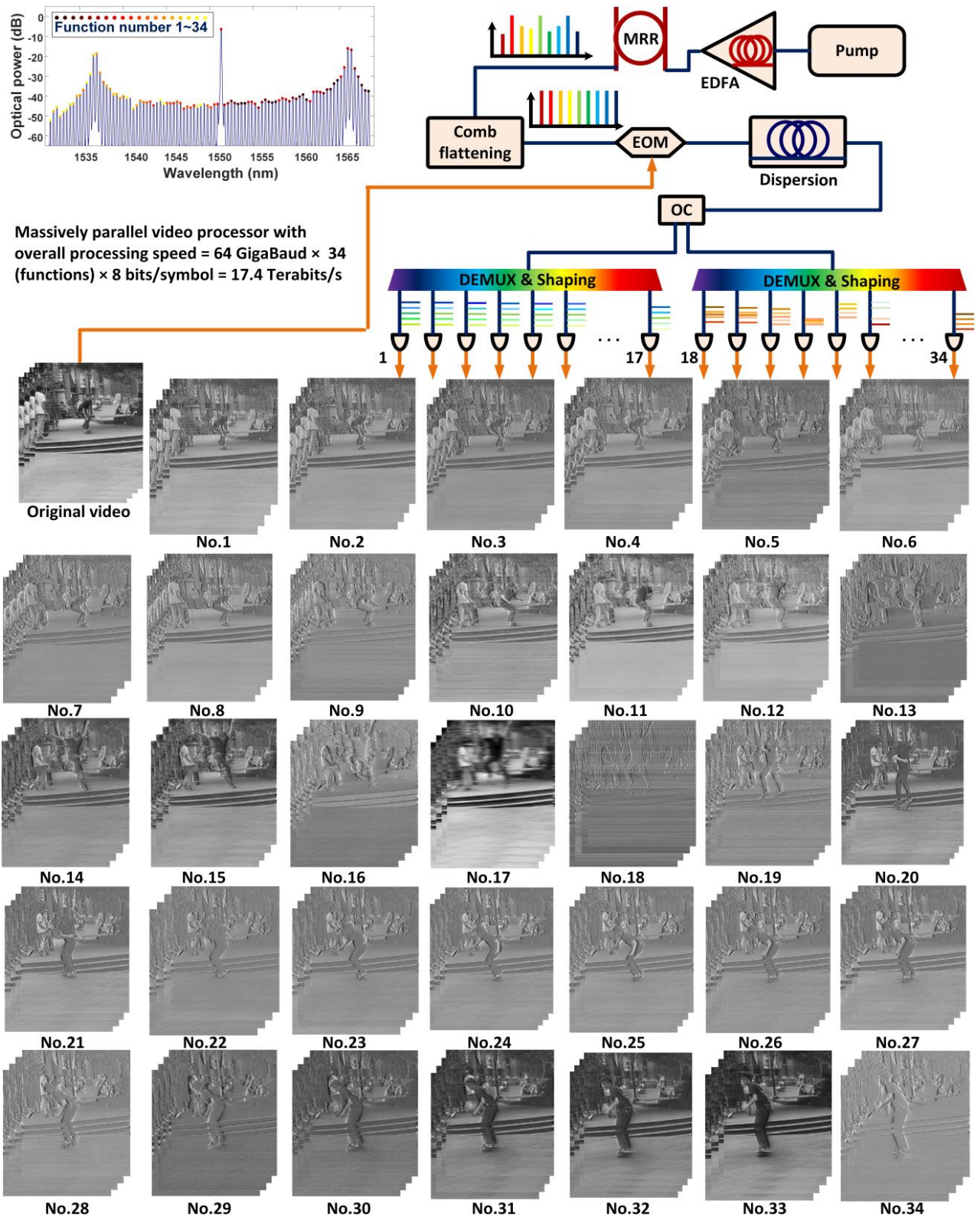

**Figure 2. Massively parallel multi-functional video processing.** EDFA: erbium doped fibre amplifier. MRR: micro-ring resonator. EOM: electro-optical Mach-Zehnder modulator. SMF: single-mode fibre. WS: WaveShaper. PD: photodetector. OC: Optical Coupler. Detailed parameters for each function have been shown in Supp. Table S1.

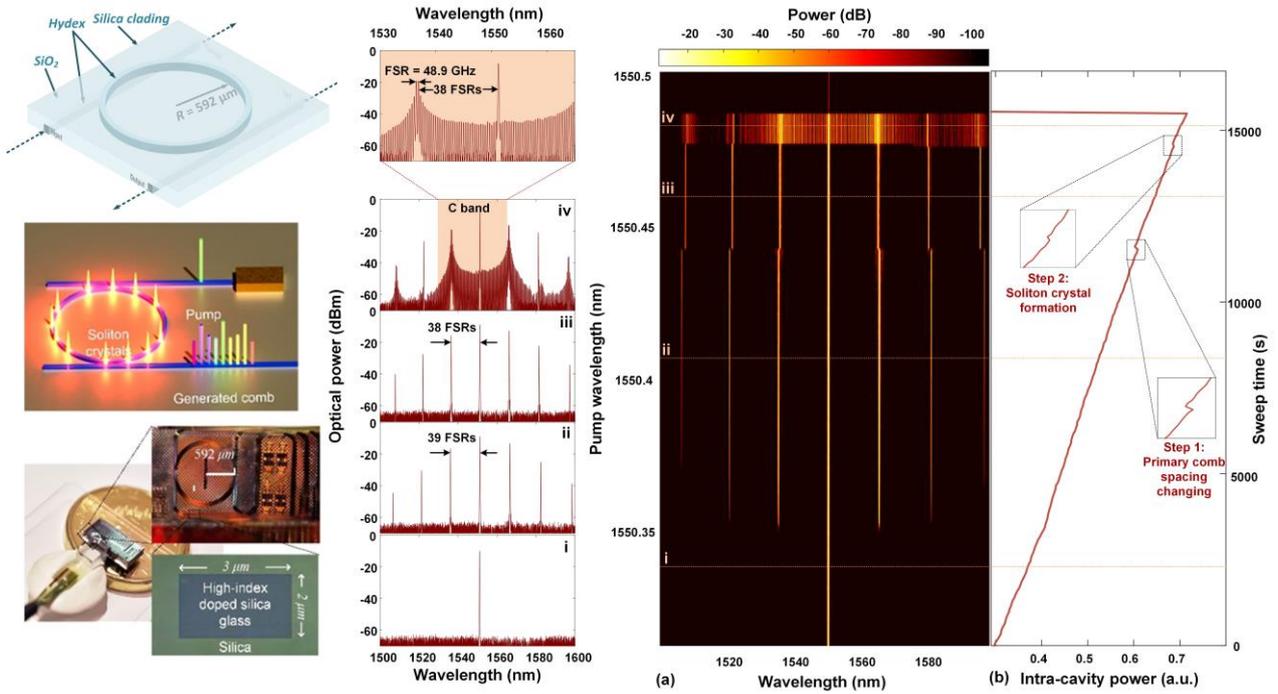

**Figure 3. Soliton crystal (SC) microcomb used for video image processing.** The SC microcomb is generated in a 4-port integrated micro-ring resonator (MRR) with an FSR of 48.9 GHz. Optical spectra of (i) Pump. (ii) Primary comb with a spacing of 39 FSRs. (iii) Primary comb with a spacing of 38 FSRs. (iv) SC micro-comb. (a) Optical spectrum of the micro-comb when sweeping the pump wavelength. (b) Measured soliton crystal step of the intra-cavity power.

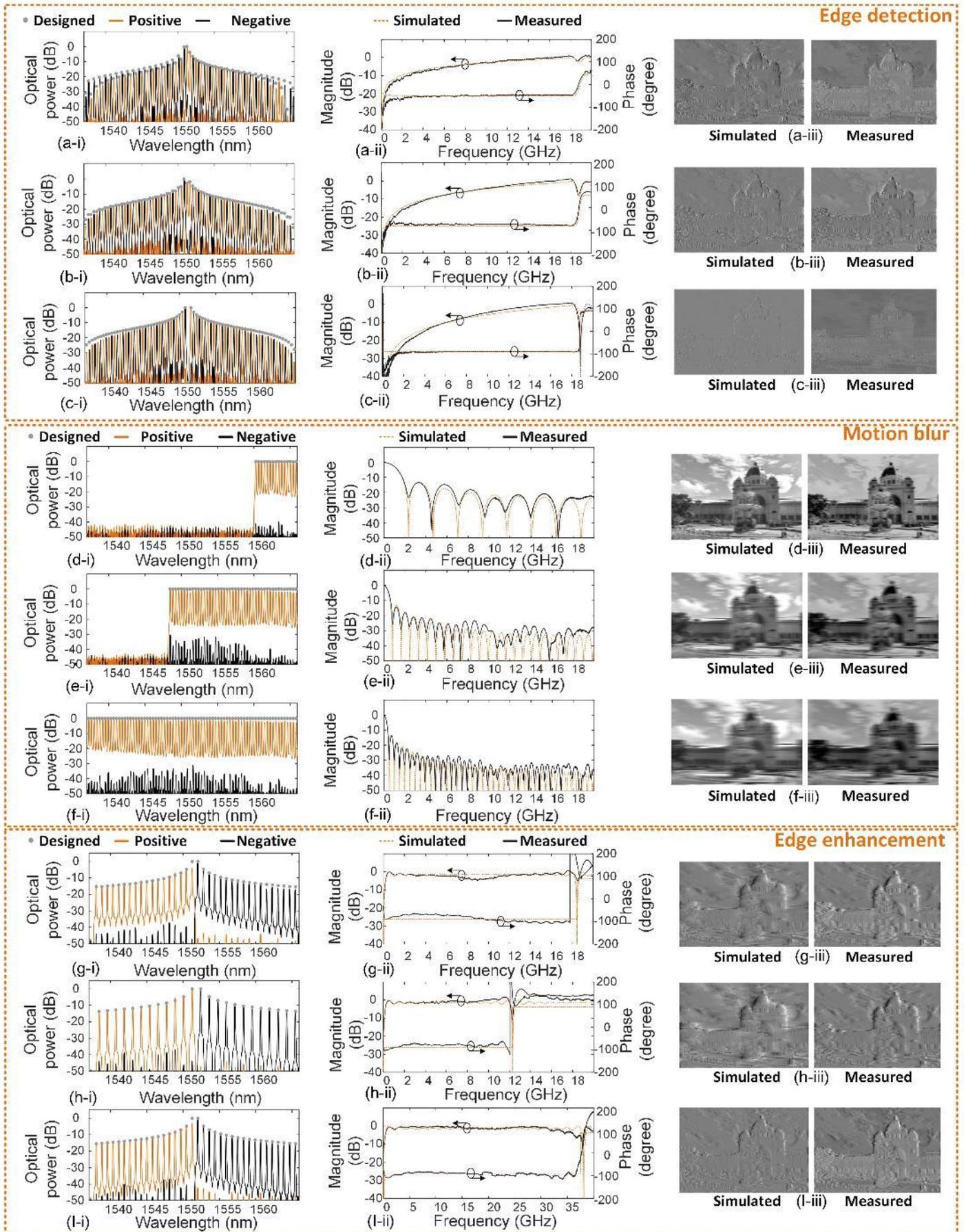

**Figure 4. Experimental results of image processing.** (a) – (c) Results for edge detection based on differentiation with order of 0.5, 0.75, and 1, respectively. (d) – (f) Results for motion blur based on integration with tap number of 15, 45, and 75, respectively. (g) – (I) Results for edge enhancement based on Hilbert transformation with operation bandwidth of 18 GHz, 12 GHz, and 38 GHz, respectively. In (a) – (I), (i) shows the designed and measured optical spectra of the shaped microcomb, (ii) shows the measured and simulated spectral response of the video image processing system, and (iii) shows the measured and simulated high definition (HD) video images after processing.

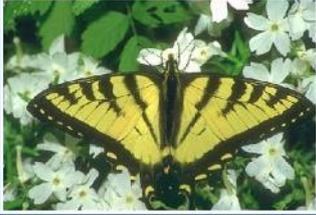

**Figure 5. Comparison of BSD images processed using the Sobel's algorithm with video image processor after edge detection.** Differentiation with different orders of 0.2, 0.4, 0.6, 0.8, and 1.0 are used for the edge detection with our video image processor. The Sobel results were performed electronically.

**Table 1. Comparison of performance parameters of images processed using Sobel's algorithm and our video image processor.**

| BSD image No. | 118035 | | 42049 | | 35010 | |
|---|---|---|---|---|---|---|
| | PR | F-Measure | PR | F-Measure | PR | F-Measure |
| Sobel | 12.2488 | 0.0068996 | 15.8362 | 0.0049214 | 11.3226 | 0.011988 |
| Differentiation order – 0.2 | 12.7891 | 0.016586 | 20.1547 | 0.021603 | 18.3944 | 0.021787 |
| Differentiation order – 0.4 | 13.629 | 0.014815 | 20.9594 | 0.024451 | 18.6818 | 0.022378 |
| Differentiation order – 0.6 | 15.249 | 0.013748 | 21.0244 | 0.024771 | 19.8858 | 0.02388 |
| Differentiation order – 0.8 | 16.2559 | 0.013491 | 21.4386 | 0.022484 | 20.0704 | 0.023491 |
| Differentiation order – 1.0 | 17.4338 | 0.014391 | 22.1003 | 0.020635 | 20.6862 | 0.026481 |